\documentclass[11pt]{article}
\usepackage{amssymb}
\usepackage{epsfig}
\begin{document}
\title{{\bf{\Large Gravitational anomalies and entropy}}}
\author{ 
 {\bf {\normalsize Bibhas Ranjan Majhi}$
$\thanks{E-mail: bibhas@iucaa.ernet.in}}\,\\
 {\normalsize IUCAA, Post Bag 4, Ganeshkhind,}
\\{\normalsize Pune University Campus, Pune 411 007, India}
\\[0.3cm]
}

\maketitle

\begin{abstract}
     A derivation of entropy from the expressions for two dimensional gravitation anomalies is given. Starting from the near horizon anomalous energy-momentum tensors corresponding to particular anomalies, the Virasoro algebra with central extension is obtained. The central charge is identified by comparing with the standard form of the algebra. Then the conserved charge in the ground state is computed. Finally, using the Cardy formula the entropy is obtained. Here both the vector and chiral theories are discussed.    
       
\end{abstract}
\vskip 9mm
\section {Introduction}
     Since the discovery of Hawking effect \cite{Hawking:1974rv}, there are several methods in the literature to understand this. The important feature is that all of them yield the universal result: {\it temperature and entropy of a black hole are proportional to the surface gravity and the horizon area, respectively} \cite{Hawking:1974rv,Bekenstein:1973ur}. Still it remains as a long standing and important question whether the black hole thermodynamics has a statistical description in terms of microscopic states. Till now there exists several ways towards this attempt: loop quantum gravity \cite{Ashtekar:1997yu}, string theory \cite{Strominger:1996sh}, the conformal field theory \cite{Brown:1986nw,Strominger:1997eq} technique, etc. But none of them is complete.

   After the discovery of the Hawking effect, people started thinking that it may play a major role to shed some light towards the quantum nature of gravity. Therefore it is necessary to have deeper  understanding of this effect. Among the existing approaches, anomaly method attracted a lot. The idea is that: it is not possible to simultaneously preserve conformal and diffeomorphism symmetries at the quantum level. In vector theory, any one of them must be sacrificed. As the latter symmetry is usually retained, there is, in general, a violation of the conformal invariance which is manifested by a non-vanishing trace of the energy-momentum tensor and leads to trace anomaly \cite{Davies}. Whereas in the chiral theory both the trace and diffeomorphism anomalies exist \cite{AlvarezGaume:1983ig,Bardeen:1984pm,Bertlmann:2000da}. Long ago, Christansen  and Fulling \cite{Christensen:1977jc} first determined the Hawking flux in the vector theory for a two dimensional black hole.

    Recently, the method has been extended to the higher dimensional black holes ($D\geq 4$). This is mainly based on the fact that near the event horizon the effective theory is two dimensional and the background is effectively the two dimensional ($t-r$) black hole metric \cite{Robinson:2005pd,Iso:2006wa,Umetsu:2009ra}. Now since, the ingoing modes are absorbed classically by the black hole, the theory must be chiral near the horizon and at the quantum level both the trace and diffeomorphism anomalies will appear. In this case, there exists two types of anomaly expressions: consistent and covariant expressions. They are related by a local counter term \cite{AlvarezGaume:1983ig,Bardeen:1984pm,Bertlmann:2000da}. Robinson and Wilczek \cite{Robinson:2005pd} showed that the Hawking flux can be obtained from the consistent anomaly expression with the use of the covariant boundary condition. In this case, both the Ward identities were required. Later on, an elegant and conceptually clearer derivation was given in \cite{Banerjee:2007qs} which was based on the covariant anomaly expression {\footnote {For more details and several important issues, see \cite{Banerjee:2007uc,Banerjee:2008wq,Banerjee:2008sn,Akhmedova:2008au}.}}. Interestingly, only one Ward identity was required.

    Although, the anomaly approach has been applied for several metrics to find the emission flux, nevertheless, a connection between these anomalies and the entropy has so far been missing in the literature. Such a investigation is necessary to have complete understanding of the role played by the anomalies in gravitational physics. In this paper, we establish this connection precisely in the case of an arbitrary dimensional static black hole.

     We briefly give our methodology to compute the entropy using the near horizon gravitational anomalies. The method, we shall follow here, is the conformal field theory (CFT) technique, particularly the work of Solodukhin \cite{Solodukhin:1998tc}. It is well known that the relevant conformal theory is the Liouville theory \cite{Coussaert:1995zp} living on the two dimensional boundary. In this theory, the conserved charges corresponding to near horizon diffeomorphism symmetry generators obey the standard Virasoro algebra with central extension and it leads to central charge \cite{Solodukhin:1998tc} {\footnote{For a Hamiltonian or Lagrangian approach and its related issues to find the Virasoro algebra, see \cite{Carlip:1998wz,Carlip:1999cy,Lin:1999gf,Das:2000zs,Jing:2000yn,Terashima:2001gn,Koga:2001vq,Park:2001zn,
Carlip:2002be,Cvitan:2002cs,Park:1999tj,Park:1999hs,Dreyer:2001py,Silva:2002jq,Brustein:2000fw,
Cvitan:2002rh,Giacomini:2003cg,Dias:2006vy,Majhi:2011ws}.}}. Then evaluating the ground state conserved charge one can find the entropy using Cardy formula \cite{Cardy:1986ie}.  
Here, we shall first cast the anomalous energy-momentum tensors corresponding to respective theories (vector and chiral theories) to the identical form of that of Liouville theory, since this is a relevant theory for the conformal field theory \cite{Coussaert:1995zp}. This will be done by using the transformations of metric and the auxiliary scalar field. Using this transformed tensor, the conserved charges $L_n$, corresponding to the near horizon diffeomorphism symmetry generators \cite{Carlip:1998wz,Carlip:1999cy}, will be defined. Computation of the commutator among $L_n$ will lead to the standard Virasoro algebra with the central extension. Then one can immediately identify the central charge. The other important quantity is the conserved charge corresponding to the zero mode, i.e. $L_0$. This will be obtained by straight forward evaluation. We shall then apply the Cardy formula \cite{Cardy:1986ie} to obtain the expression for entropy. Generally two copies of Virasoro algebra, one for $L_n$ (holomorphic) and other for ${\bar{L}}_n$ (anti-holomorphic), contribute to the entropy. However, in our case since the theory is near horizon where only ingoing modes contribute, and so only one copy, namely $L_n$, is sufficient to calculate the entropy.

   The organization of the paper is as follows. In the next section, the near horizon effective two dimensional metric will be introduced. Here it will be written in the null coordinates which will be relevant for our subsequent analysis. In section 3, the energy-momentum tensor in two dimensional vector theory will be casted to that of Liouville theory. Defining the conserved charge, we will obtain the Virasoro algebra. Then by the Cardy formula the entropy will be obtained. Next section will be devoted for the same in the case of chiral theory. Final section will contain the conclusions.         

\section{Metric and null coordinates}
     The physics of a black hole is mainly dominated by the event horizon properties. Near this horizon, the theory is effectively determined by the two dimensional ($t-r$) metric \cite{Robinson:2005pd,Iso:2006wa,Umetsu:2009ra}:
\begin{eqnarray}
ds^2 = F(r)dt^2 - \frac{dr^2}{F(r)}
\label{1.08}
\end{eqnarray}
whose event horizon is defined by the relation $F(r=r_H) = 0$ and the surface gravity is given by $\kappa = \frac{F'(r_H)}{2}$.
For our later analysis, it will be appropriate to use the null tortoise coordinates, defined as, 
\begin{eqnarray}
u = t-r^*, \,\,\ v = t+r^*; \,\,\,\ dr^* = \frac{dr}{F(r)}~.
\label{1.09}
\end{eqnarray}
In these coordinates metric (\ref{1.08}) takes to the following form
\begin{eqnarray}
ds^2 = \frac{F(r)}{2} (dudv + dvdu)~.
\label{1.10}
\end{eqnarray}

  The justification of choosing the effective $2D$ metric (\ref{1.08}) for our purpose of finding entropy by two dimensional gravitational anomalies in the context of Virasoro algebra is as follows. 
Consider first the following four dimensional metric:
\begin{eqnarray}
ds^2_4 = F(r)dt^2 - \frac{dr^2}{F(r)} - r^2(d\theta^2 + \sin^2\theta d\phi^2)~.
\label{metric1}
\end{eqnarray}
The horizon (apparent) is defined as the curve $H$ such that \cite{Russo:1995zm},
\begin{eqnarray}
g^{ab}\nabla_a r\nabla_b r|_H = 0~,
\label{metric2}
\end{eqnarray}    
where $g_{ab}$ is the ($t-r$) sector of the full metric (\ref{metric1}). This clearly says $F(r=r_H)=0$, as stated earlier. Now the above condition is invariant under conformal transformation $g_{ab}\rightarrow \Omega^2 g_{ab}$ and so the two dimensional metric is determined by this condition only upto a conformal factor. Interestingly, very near to the horizon, the diffeomorphisms which preserve the condition (\ref{metric2}), form the infinite dimensional group of conformal transformations in two dimensions. The corresponding generators satisfy Virasoro algebra. Also, the fields become massless. For details, see section 2 of \cite{Solodukhin:1998tc}. Therefore, the physics, near the horizon, is conformal and governed by an effective two dimensional field theory. Thus we see that the horizon defining condition (\ref{metric2}) is essentially a condition on the radial function $r$ of the full metric (\ref{metric1}) and the two dimensional metric is determined upto a conformal factor. So the horizon dynamics is encoded in the radial function while the $2$D metric is chosen from the same conformal class. Therefore, here we have chosen the two dimensional metric as (\ref{1.08}) and it is feasible to study conformal field theory on this background.

\section{Vector theory, Virasoro algebra and entropy}
    In this section, the entropy will be evaluated from the anomalous stress tensor for the vector theory. 
The idea is that near the event horizon, the theory is effectively dominated by the two dimensional metric of the form (\ref{1.10}). For details, see \cite{Robinson:2005pd,Iso:2006wa,Umetsu:2009ra}. Now the energy-momentum tensor for the scalar field under this effective two dimensional metric, at the quantum level, leads to non-vanishing trace since usually the diffeomorphism symmetry is retained. Such theory is called as vector theory.

  In this case, the trace anomaly is given by \cite{Deser:1976yx}
\begin{eqnarray}
\tilde{T}{^a_a} = \frac{\tilde R}{24\pi}~,
\label{1.01}
\end{eqnarray}
and the energy-momentum tensor satisfies the conservation equation  $\tilde{\nabla}^a\tilde T_{ab} = 0$. 
The anomaly induced effective action $S_P$, obtained from the functional integration of $\tilde{T^a_a} = \frac{2}{\sqrt{-\tilde{g}}}\tilde{g}^{ab}\frac{\delta S_P}{\delta\tilde{g}^{ab}}$, is given by the Polyakov action \cite{Polyakov:1981rd},
\begin{eqnarray}
S_P = -\frac{1}{96\pi}\int d^2x\sqrt{-{\tilde{g}}} {\tilde{R}}\frac{1}{\tilde{\Box}}{\tilde{R}}~.
\label{effective1}
\end{eqnarray} 
This action is non-local and it can be written in local form by introducing an auxiliary field $\tilde{\Phi}$:
\begin{eqnarray}
S_P = -\frac{1}{96\pi}\int d^2x{\sqrt{-\tilde{g}}} (-\tilde{\Phi} \tilde{\Box} \tilde{\Phi} + 2\tilde{\Phi} \tilde{R})~.
\label{effective2}
\end{eqnarray}
The two dimensional anomalous energy-momentum tensor is given by, 
\begin{eqnarray}
{\tilde{T}}_{ab}= \frac{2}{\sqrt{-\tilde{g}}}\frac{\delta S_P}{\delta\tilde{g}^{ab}} = \frac{1}{48\pi}\Big[\tilde{\nabla}_a\tilde{\Phi}\tilde{\nabla}_b\tilde{\Phi} - 2\tilde{\nabla}_a\tilde{\nabla}_b\tilde{\Phi} + \tilde{g}_{ab}\Big(2\tilde{R} - \frac{1}{2}\tilde{\nabla}_c\tilde{\Phi}\tilde{\nabla}^c\tilde{\Phi}\Big)\Big]
\label{1.02}
\end{eqnarray} 
with the auxiliary scalar $\tilde{\Phi}$ satisfies the equation of motion,
\begin{eqnarray}
\tilde{\Box} \tilde{\Phi} = \tilde{R}~.
\label{1.03}
\end{eqnarray}
Here $\tilde R$ is the two dimensional Ricci scalar. 
Christensen and Fulling \cite{Christensen:1977jc} evaluated the Hawking flux by solving the conservation equation and the trace anomaly expression (\ref{1.01}) for the Schwarzschild black hole. Here, in the following, we will find the entropy from the expression for anomalous energy-momentum tensor (\ref{1.02}).

   Now it is known that the Liouville theory is conformally invariant and possesses two sets of Virasoro generators $L_n$ and $\bar{L}_n$ \cite{Coussaert:1995zp}. Therefore, it is a relevant CFT and so it is convenient to recast the energy-momentum tensor (\ref{1.02}) into that of the Liouville theory so that it fits into the Virasoro scheme. The same technique was also adopted in \cite{Solodukhin:1998tc}, where the four dimensional Einstein-Hilbert action was casted to Liouville action by dimensional reduction technique and then imposing a set of particular transformations. Motivated by this fact, here also we will do the same. This is done by the following set of transformations, 
\begin{eqnarray}
\tilde{g}_{ab} =  e^{\Big(\frac{2}{qr_H} - \frac{qr_H}{16}\Big)\Phi} g_{ab} ; \,\,\,\ \tilde{\Phi} = \frac{qr_H}{8}\Phi~.
\label{1.04}
\end{eqnarray}
Under these, (\ref{1.02}) and (\ref{1.03}) transform to 
\begin{eqnarray}
&&T_{ab} = \frac{1}{48\pi}\Big[\frac{1}{2}\nabla_a\Phi\nabla_b\Phi - \frac{qr_H}{4}\nabla_a\nabla_b\Phi + g_{ab}\Big(2R - \frac{1}{4}\nabla_c\Phi\nabla^c\Phi - \frac{4}{qr_H}\nabla_c\nabla^c\Phi 
\nonumber
\\
&&+ \frac{qr_H}{8}\nabla_c\nabla^c\Phi\Big)\Big]
\label{1.05}
\\
&&\Box\Phi = \Big(\frac{qr_H}{16} + \frac{2}{qr_H}\Big)^{-1} R
\label{1.06}
\end{eqnarray}
where $r_H$ is the radius of the event horizon and $q$ is a constant.
Substituting (\ref{1.06}) in (\ref{1.05}), the energy-momentum tensor reduces to the following form
\begin{eqnarray}
T_{ab} = \frac{1}{48\pi}\Big[\frac{1}{2}\nabla_a\Phi\nabla_b\Phi - \frac{qr_H}{4}\nabla_a\nabla_b\Phi + g_{ab}\Big(\frac{qr_H}{4}\nabla_c\nabla^c\Phi - \frac{1}{4}\nabla_c\Phi\nabla^c\Phi\Big)\Big]~.
\label{1.07}
\end{eqnarray}
This is similar to the energy-momentum tensor of Liouville theory as obtained by Solodukhin \cite{Solodukhin:1998tc} from the four dimensional Einstein-Hilbert action. Here, we obtained this from the two dimensional trace anomaly expression by applying the transformations (\ref{1.04}).

      Before proceeding further, note that because of non-vanishing of trace 
\begin{eqnarray}
T = \frac{1}{48\pi} \frac{qr_H}{4}\Box\Phi~,
\label{trace1}
\end{eqnarray}
corresponding to the stress tensor (\ref{1.07}), the theory is not conformal. However, within the small vicinity of the horizon, we will show below that the theory becomes conformal. In the ($t,r^*$) coordinate, equation of motion for $\Phi$ (\ref{1.06}) under the background (\ref{1.08}) reads,
\begin{eqnarray}
- \partial^2_t\Phi + \partial^2_{r^*}\Phi = F \Big(\frac{qr_H}{16} + \frac{2}{qr_H}\Big)^{-1} R~.
\label{trace2}
\end{eqnarray}
Now for the metric (\ref{1.08}), $R$ is non-zero and since $F\rightarrow 0$ near the horizon, right hand side of the above is negligible. Therefore, (\ref{trace2}) reduces to 
\begin{eqnarray}
\partial^2_t\Phi - \partial^2_{r^*}\Phi = 0~,
\label{trace3}
\end{eqnarray}
which signifies that due to existence of the horizon, the free scalar field is propagating in flat spacetime near the horizon. Hence the trace (\ref{trace1}) vanishes and it implies that the Poisson algebra of the components of trace tensor (\ref{1.07}) closes and they form the Virasoro algebra very near to the horizon. Therefore, the theory of the scalar field $\Phi$, whose tress tensor is given by (\ref{1.07}), is conformal when one considers it in the {\it small vicinity} of the horizon.

   Let us now proceed to obtain the Virasoro algebra among the charges corresponding to the near horizon diffeomorphism symmetry generators. Since near the horizon the ingoing modes contribute, only the $T_{vv}$ component of (\ref{1.07}) is important. Under the metric (\ref{1.10}), $T_{vv}$ component comes out to be
\begin{eqnarray}
T_{vv} = \frac{1}{48\pi} \Big[\frac{1}{2}\partial_v\Phi\partial_v\Phi - \frac{qr_H}{4}\Big(\partial_v^2\Phi - \frac{\partial_v F(r)}{F(r)}\partial_v\Phi\Big)\Big]~.
\label{1.11}
\end{eqnarray}
Again, since we are interested only near the horizon where $\frac{\partial_v F(r)}{F(r)} \rightarrow \frac{F'(r_H)}{2}=\kappa$, the above reduces to the following form,
\begin{eqnarray}
T_{vv} = \frac{1}{48\pi} \Big[\frac{1}{2}\partial_v\Phi\partial_v\Phi - \frac{qr_H}{4}\partial_v^2\Phi + \frac{qr_H\kappa}{4}\partial_v\Phi\Big]~.
\label{1.12}
\end{eqnarray}

   Here, we shall determine the entropy  by counting the horizon microstates of (\ref{1.08}) via Cardy formula \cite{Cardy:1986ie},
\begin{eqnarray}
S = 2\pi \sqrt{\frac{CL_0}{6}}
\label{1.13}
\end{eqnarray}
corresponding to the quantum Virasoro algebra
\begin{eqnarray}
[L_n,L_m] = (n-m)L_{n+m} + \frac{C}{12}n(n^2 - 1)\delta_{n+m,0}~.
\label{1.14}
\end{eqnarray}
In the above, $[,]$ is the commutator while $C$ and $L_n$ are the central charge and the conserved charge for $n$th mode respectively. For the present case we define $L_n$ as,
\begin{eqnarray}
L_n = \int dv \xi_n(v)T_{vv}(v)
\label{1.15}
\end{eqnarray}
where $\xi_n(v)$ is the near horizon diffeomophism symmetry generator, obeying one sub-algebra isomorphic to Diff $S^1$,
\begin{eqnarray}
[\xi_n,\xi_m] = (n-m)\xi_{n+m}~.
\label{1.16}
\end{eqnarray}    
Following \cite{Carlip:1999cy}, the explicit form of these generators for the metric (\ref{1.08}) satisfying (\ref{1.16}) can be obtained. These are given by,
\begin{eqnarray}
\xi_n = \frac{1}{\kappa} e^{in\kappa v}
\label{1.17}
\end{eqnarray}
where $n$ is an integer, can be positive and negative as well. The coordinate $v$ is periodic within the range $0\leq v \leq \frac{2\pi}{\kappa}$ and hence we choose this as the limits of the integration (\ref{1.15}).

   Now, in order to obtain the commutator of the conserved charges (\ref{1.15}), let us first calculate the Poisson bracket $\{L_n,L_m\}$. This will contain a Poisson bracket $\{T_{vv}(v),T_{vv}(\bar{v})\}$. To evaluate the bracket, we need the following basic Poisson bracket relation 
\begin{eqnarray}
\{\Phi(r^*,t), \partial_t\Phi(\bar{r^*},t)\} = 48\pi\delta (r^*-\bar{r^*})
\label{1.18}
\end{eqnarray}
and so consequently, we have,
\begin{eqnarray}
\{(\partial_t + \partial_{r^*})\Phi(r^*,t), (\partial_t+\partial_{\bar{r^*}})\partial_t\Phi(\bar{r^*},t)\} = 96\pi\partial_{r^*}\delta (r^*-\bar{r^*})~.
\label{1.19}
\end{eqnarray}
The bracket (\ref{1.18}) is actually the bracket among the field and its conjugate momentum. The normalization in the right hand side has been fixed by the normalization in the corresponding action.  
In radial null coordinates, above bracket reduces to the following form,
\begin{eqnarray}
\{\partial_v\Phi(v),\partial_{\bar{v}}\Phi(\bar{v})\}=48\pi\partial_v\delta(v-\bar{v})~.
\label{1.20}
\end{eqnarray} 
For the present case $T_{vv}(v)$ is given by (\ref{1.12}). So using (\ref{1.20}) we obtain
\begin{eqnarray}
\{T_{vv}(v), T_{vv}(\bar{v})\} &=& \frac{1}{48\pi}\Big[\partial_v\Phi(v)\partial_{\bar{v}}\Phi(\bar{v})\partial_v\delta(v-\bar{v}) - \frac{qr_H}{4}\partial_v\Phi(v)\partial_v\partial_{\bar v}\delta(v-\bar v) 
\nonumber
\\
&+& \frac{qr_H\kappa}{4} \partial_v\Phi(v)\partial_v\delta(v-\bar v) - \frac{qr_H}{4}\partial_{\bar v}\Phi(\bar v)\partial_v^2\delta(v-\bar v)
\nonumber
\\
&+& \Big(\frac{qr_H}{4}\Big)^2\partial_{\bar v}\partial_v^2\delta(v-\bar v) - \Big(\frac{qr_H}{4}\Big)^2 \kappa \partial_v^2\delta(v-\bar v)
\nonumber
\\
&+& \frac{qr_H\kappa}{4}\partial_{\bar v}\Phi(\bar v)\partial_v\delta(v-\bar v) - \Big(\frac{qr_H}{4}\Big)^2\kappa \partial_v\partial_{\bar v}\delta(v-\bar v) 
\nonumber
\\
&+& \Big(\frac{qr_H}{4}\Big)^2 \kappa^2 \partial_v\delta(v-\bar v)\Big]~. 
\label{1.21}
\end{eqnarray}
Hence, considering that the auxiliary field $\Phi$ and its derivatives with respect to $v$ are periodic within the interval $0\leq v \leq 2\pi/\kappa$, we find, 
\begin{eqnarray}
\{L_n,L_m\} &=& \int dv\Big(\xi_n\partial_v\xi_m - \xi_m\partial_v\xi_n\Big)T_{vv}-\frac{1}{48\pi}\Big(\frac{qr_H\kappa}{4}\Big)^2\int dv\xi_m\partial_v\xi_n
\nonumber
\\
&-& \frac{1}{48\pi}\Big(\frac{qr_H}{4}\Big)^2 \int dv\partial_v\xi_m\partial_v^2\xi_n~.
\label{1.22}
\end{eqnarray}
Finally, substituting the expressions for $\xi_n$ from (\ref{1.17}) in the above and performing the integration we obtain,
\begin{eqnarray}
i\{L_n,L_m\} = (n-m)L_{n+m} + \frac{1}{24} \Big(\frac{qr_H}{4}\Big)^2 (n^3+n) \delta_{n+m,0}~.
\label{1.23}
\end{eqnarray}
Application of the usual prescription for getting the commutator relation from the Poisson bracket yields, in the unit $\hbar=1$,
\begin{eqnarray}
[L_n,L_m] = i\{L_n,L_m\} = (n-m)L_{n+m} + \frac{1}{24} \Big(\frac{qr_H}{4}\Big)^2 (n^3+n) \delta_{n+m,0}~.
\label{1.23new}
\end{eqnarray}
In order to put this similar to the standard form (\ref{1.14}), let us redefine $L_n$ as,
\begin{eqnarray}
L_n \rightarrow L_n - \frac{1}{24} \Big(\frac{qr_H}{4}\Big)^2 \delta_{n,0}~.
\label{1.24}
\end{eqnarray}
This redefinition leads to the following algebra:
\begin{eqnarray}
[L_n,L_m] = (n-m)L_{n+m} + \frac{1}{24} \Big(\frac{qr_H}{4}\Big)^2 n(n^2-1) \delta_{n+m,0}~.
\label{1.25}
\end{eqnarray}
Comparing this with (\ref{1.14}) we obtain the central charge for the black hole case,
\begin{eqnarray}
C= \frac{q^2r_H^2}{32}~.
\label{1.26}
\end{eqnarray}

   Next step is to find the value of $L_0$. By the definition for $L_n$ (\ref{1.15}) we have,
\begin{eqnarray}
L_0 = \frac{1}{48\pi\kappa}\int dv  \Big[\frac{1}{2}\partial_v\Phi\partial_v\Phi - \frac{qr_H}{4}\partial_v^2\Phi + \frac{qr_H\kappa}{4}\partial_v\Phi\Big]~.
\label{1.27}
\end{eqnarray}
Here, since $\Phi$ and its derivative terms with respective to coordinate $v$ are periodic within the interval $0\leq v\leq 2\pi/\kappa$, the following intregal vanishes: 
\begin{eqnarray}
&&\int_0^{2\pi/\kappa} dv~\partial_v^2\Phi = \partial_v\Phi|_0^{2\pi/\kappa} = 0,
\nonumber
\\
&&\int_0^{2\pi/\kappa} dv~\partial_v\Phi = \Phi|_0^{2\pi/\kappa} = 0~.
\label{1.28}
\end{eqnarray}
To evaluate the first integral in (\ref{1.27}) we will use the following trick. Let us first find the solution of $\Phi$ from (\ref{1.06}). For simplicity we consider the Schwarzschild black hole. Also, our region of interest is near the event horizon and therefore we need only ($t-r$)-sector of the metric. In this case, the solution is given by \cite{Balbinot:1999vg},
\begin{eqnarray}
\Phi = \Big(\frac{qr_H}{16} + \frac{2}{qr_H}\Big)^{-1} \Big[at - \ln \Big(1-\frac{2M}{r}\Big) + A \Big\{r+2M\ln \Big(r-2M\Big)\Big\}+B\Big]~,
\label{1.28new1}
\end{eqnarray}
where $a$, $A$ and $B$ are arbitrary constants. Now it has been shown that the relevant vacuum state to obtain the Hawking flux, in the context of gravitational anomalies (both in vector case \cite{Balbinot:1999vg} and in chiral case \cite{Banerjee:2008wq}), is the Unruh vacuum. In this vacuum, the solution for $\Phi$ near the horizon comes out to be \cite{Balbinot:1999vg},
\begin{eqnarray}
\Phi \approx -\Big(\frac{qr_H}{16} + \frac{2}{qr_H}\Big)^{-1}  \frac{v}{4M} + \textrm{constant}~.
\label{1.28new2}
\end{eqnarray}
Using this value we obtain,
\begin{eqnarray}
\int_0^{2\pi/\kappa} dv ~(\partial_v\Phi)^2 &=& -\int_0^{2\pi/\kappa} dv ~\Phi\partial^2\Phi \approx \Big(\frac{qr_H}{16} + \frac{2}{qr_H}\Big)^{-1}\frac{1}{4M} \int_0^{2\pi/\kappa}dv~v\partial^2_v\Phi
\nonumber
\\
&=& -\Big(\frac{qr_H}{16} + \frac{2}{qr_H}\Big)^{-1}\frac{1}{4M} \int_0^{2\pi/\kappa}dv~\partial_v\Phi =0~,
\label{1.28new}
\end{eqnarray}
and so $L_0$ vanishes. Therefore, the redefined $L_n$ (\ref{1.24}) yields,
\begin{eqnarray}
L_0 =\frac{q^2r_H^2}{384} 
\label{1.29}
\end{eqnarray}
Now, substituting (\ref{1.26}) and (\ref{1.29}) in the Cardy formula (\ref{1.13}), we obtain the entropy,
\begin{eqnarray}
S = \frac{q^2l_p^2}{96\sqrt{2}} S_{BH}
\label{1.30}
\end{eqnarray}
where $S_{BH} = \frac{\pi r_H^2}{l_p^2}$ is the Bekenstein-Hawking entropy and $l_p$ is the Planck length. Note that the conformal entropy $S$ is equal to the Bekenstein-Hawking entropy if we choose $q^2 = \frac{96\sqrt{2}}{l_p^2}$. 

   Now several comments are in order. (i) To see that there is no dimensional ambiguity in the  transformations (\ref{1.04}), remember, in two dimensions $\Phi$ is dimensionless if we choose Lorentz-Heaviside unit. Hence $qr_H$ must be dimensionless and this is actually the case since the dimension of $q$, as we just observed, is inverse of length. (ii) Here we obtained the entropy from the anomalies which are the result of the quantization of the matter fields near the horizon. So the entropy (\ref{1.30}) is not the entropy of the black hole, rather it can be interpreted as that of the matter. But interestingly, the matter entropy is proportional to the horizon area. Similar has been observed earlier in \cite{Kolekar:2010py} for a freely falling box of ideal gas in a black hole. (iii) Both the central charge $C$ and the zero mode eigenvalue $L_0$ depend on $q$. The dependence of $C$ on $q$ is similar to \cite{Solodukhin:1998tc} while that of $L_0$ on $q$ is different. This is because the we applied a different method to find $L_0$. Solodukhin determined it from a more general zero mode solution of field equation near the horizon and imposing specific boundary conditions. While we found it by just redefining $L_n$ and hence in our case $L_0$ is proportional to $q^2$. Therefore the final expression for entropy contains the parameter $q$. This ambiguity is because a direct evaluation leads to vanishing $L_0$ for the classical configuration. Depending on the regularization of the ground state one can obtain a non-zero value and different regularization leads to different value. Of course, one might be interested to see if there exists any method in which $L_0$ will come out to be inversely proportional to $q^2$ and in that case the expression for entropy will not contain any undetermined coefficient.

\section{Chiral theory, Virasoro algebra and entropy}
    In chiral theory, if one of the symmetries (conformal or diffeomorphism) breaks down then naturally other must breaks down. Hence both trace and diffeomorphism anomalies will appear in this theory. The analogous energy-momentum tensor is given by \cite{Leutwyler:1984nd,Banerjee:2007uc,Banerjee:2008wq},
\begin{eqnarray}
\tilde{T}_{ab} = \frac{1}{48\pi} \Big[\frac{1}{4}\tilde{D}_a\tilde{G}\tilde{D}_b\tilde{G} - \frac{1}{2}\tilde{D}_a\tilde{D}_b\tilde{G} + \frac{1}{2}\tilde{g}_{ab}\tilde{R}\Big]
\label{1.31}
\end{eqnarray}  
where $\tilde{D}_a = \tilde{\nabla}_a\pm \tilde{\bar\epsilon}_{ab}\tilde{\nabla}^b$ is the chiral derivative and the auxiliary field $\tilde{G}$ satisfies the equation of motion,
\begin{eqnarray}
\tilde{\Box}\tilde{G} = \tilde{R}~.
\label{1.32}
\end{eqnarray}
Here $+$ ($-$) corresponds to the ingoing (outgoing) mode. Since, again the near horizon theory will be considered where only ingoing mode exists, we will concentrate on the plus sign. This stress tensor corresponds to both the trace and diffeomorphism anomalies which are in covariant form. The expressions are:
\begin{eqnarray}
\tilde{T}^a_a = \frac{\tilde R}{48\pi};\,\,\,\ \tilde{\nabla}^a\tilde{T}_{ab} = - \frac{1}{96\pi}\tilde{\bar{\epsilon}}_{bc}\tilde{\nabla}^c\tilde{R}~.
\label{chiral1}
\end{eqnarray}
   
  Now, as earlier, to get a form of energy-momentum tensor similar to that of Liouville theory the transformations (\ref{1.04}) will be applied on (\ref{1.31}) and (\ref{1.32}). Then proceeding in the identical way we obtain the following transformed energy-momentum tensor,
\begin{eqnarray}
T_{ab} = \frac{1}{48\pi}\Big[\frac{1}{8}D_aG D_bG -\frac{qr_H}{16}D_aD_b G +\frac{1}{4}g_{ab}\Big(\frac{qr_H}{4}\Box G - \frac{1}{4}\nabla_cG\nabla^cG + \frac{1}{2}(\frac{qr_H}{8})^2 \nabla_cG\nabla^cG\Big)\Big]~.
\label{1.33}
\end{eqnarray}
Near the horizon, since only ingoing modes will be considered, under the background metric (\ref{1.10}) we must have,
\begin{eqnarray}
D_v = 2\nabla_v;\,\,\,\ D_u = 0~.
\label{1.34}
\end{eqnarray} 
Therefore, $T_{vv}$ component is given by (\ref{1.11}). Hence following the identical steps we find the entropy as (\ref{1.30}).

\section{Conclusions}
    The effective theory near the event horizon of black hole corresponds to ($1+1$)-dimensional metric and is conformal. Such situation gives rise to the two dimensional gravitational anomaly at the quantum level. Depending on the nature of the theory (vector or chiral), there exits trace \cite{Davies} or both trace and diffeomorphism anomalies \cite{AlvarezGaume:1983ig,Bardeen:1984pm,Bertlmann:2000da}. It has been shown that the Hawking flux can be derived from these anomaly expressions \cite{Christensen:1977jc,Robinson:2005pd,Iso:2006wa,Banerjee:2007qs}. So it may be possible that the anomalies can play a crucial role to illuminate the quantum nature of a black hole. Hence, one has to see if they have a wide applicability in the different aspects of the gravity.

   In this paper, we made an attempt to the thermodynamics of the black hole system. Here we found the entropy from the anomalous energy-momentum tensor corresponding to the particular anomaly/anomalies. We discussed both vector as well as chiral theories. We adopted the conformal field theory technique proposed earlier in \cite{Brown:1986nw,Strominger:1997eq,Solodukhin:1998tc}. First, the expressions for energy-momentum tensors were casted to that of Liouville theory, since this a relevant conformal theory. Then the conserved charges corresponding to the near horizon ingoing tensor and diffeomorphism symmetry generators were defined. An explicit calculation of commutator between these charges gave rise to the Virasoro algebra which had the central extension. Finally, identifying the central charge and calculating the charge of the zero mode, we derived the entropy by the Cardy formula.

   It was noted that the entropy came out to be proportional to the well known Bekenstein-Hawking expression. The proportionality constant contained a undetermined parameter ``$q$'', originally appeared in the transformations (\ref{1.04}). To obtain the exact expression one needed to choose a particular value of it. The similar situation has been encountered earlier in \cite{Cvitan:2002rh,Dias:2006vy}. The independent derivation of such choice may be interesting and will make this method more appealing. Unfortunately, it is not known to the author. 

   Despite this limitation, our paper addressed a glaring omission in the literature. A way to find the entropy from the gravitational anomalies has been given. This may explore the importance of anomalies in studying the quantum nature of gravity.


\begin{thebibliography}{99}
\bibitem{Hawking:1974rv}
  S.~W.~Hawking,
  Nature {\bf 248}, 30 (1974).
\bibitem{Bekenstein:1973ur}
  J.~D.~Bekenstein,
  Phys.\ Rev.\  D {\bf 7}, 2333 (1973).
\bibitem{Ashtekar:1997yu}
  A.~Ashtekar, J.~Baez, A.~Corichi and K.~Krasnov,
  Phys.\ Rev.\ Lett.\  {\bf 80}, 904 (1998)
  [arXiv:gr-qc/9710007].\\
  J.~M.~Garcia-Islas,
  Class.\ Quant.\ Grav.\  {\bf 25}, 245001 (2008)
  [arXiv:0804.2082 [gr-qc]].
\bibitem{Strominger:1996sh}
  A.~Strominger and C.~Vafa,
  Phys.\ Lett.\  B {\bf 379}, 99 (1996)
  [arXiv:hep-th/9601029].
\bibitem{Brown:1986nw}
  J.~D.~Brown and M.~Henneaux,
  Commun.\ Math.\ Phys.\  {\bf 104}, 207 (1986).
\bibitem{Strominger:1997eq}
  A.~Strominger,
  JHEP {\bf 9802}, 009 (1998)
  [arXiv:hep-th/9712251].
\bibitem{Davies}
 P.~C.~W.~Davies and S.~A.~Fulling,
 Proc. R. Soc. Lond. {\bf A 354}, 59 (1977).
\bibitem{AlvarezGaume:1983ig}
  L.~Alvarez-Gaume, E.~Witten,
  Nucl.\ Phys.\  {\bf B234}, 269 (1984).
\bibitem{Bardeen:1984pm}
  W.~A.~Bardeen and B.~Zumino,
  Nucl.\ Phys.\  B {\bf 244}, 421 (1984).
\bibitem{Bertlmann:2000da}
  R.~A.~Bertlmann and E.~Kohlprath,
  Annals Phys.\  {\bf 288}, 137 (2001)
  [arXiv:hep-th/0011067].
\bibitem{Christensen:1977jc}
  S.~M.~Christensen and S.~A.~Fulling,
  Phys.\ Rev.\  D {\bf 15}, 2088 (1977).
\bibitem{Robinson:2005pd}
  S.~P.~Robinson and F.~Wilczek,
  Phys.\ Rev.\ Lett.\  {\bf 95}, 011303 (2005)
  [arXiv:gr-qc/0502074].
\bibitem{Iso:2006wa}
  S.~Iso, H.~Umetsu and F.~Wilczek,
  Phys.\ Rev.\ Lett.\  {\bf 96}, 151302 (2006)
  [arXiv:hep-th/0602146].
\bibitem{Umetsu:2009ra}
  K.~Umetsu,
  Int.\ J.\ Mod.\ Phys.\  A {\bf 25}, 4123 (2010)
  [arXiv:0907.1420 [hep-th]].
\bibitem{Banerjee:2007qs}
  R.~Banerjee and S.~Kulkarni,
  Phys.\ Rev.\  D {\bf 77}, 024018 (2008)
  [arXiv:0707.2449 [hep-th]].
\bibitem{Banerjee:2007uc}
  R.~Banerjee and S.~Kulkarni,
  Phys.\ Lett.\  B {\bf 659}, 827 (2008)
  [arXiv:0709.3916 [hep-th]].
\bibitem{Banerjee:2008wq}
  R.~Banerjee and S.~Kulkarni,
  Phys.\ Rev.\  D {\bf 79}, 084035 (2009)
  [arXiv:0810.5683 [hep-th]].
\bibitem{Banerjee:2008sn}
  R.~Banerjee and B.~R.~Majhi,
  Phys.\ Rev.\  D {\bf 79}, 064024 (2009)
  [arXiv:0812.0497 [hep-th]].
\bibitem{Akhmedova:2008au}
  V.~Akhmedova, T.~Pilling, A.~de Gill and D.~Singleton,
  Phys.\ Lett.\ B {\bf 673}, 227 (2009)
  [arXiv:0808.3413 [hep-th]].\\
  A.~Zampeli, D.~Singleton and E.~C.~Vagenas,
  JHEP {\bf 1206}, 097 (2012)
  [arXiv:1206.0879 [gr-qc]].
\bibitem{Coussaert:1995zp}
  O.~Coussaert, M.~Henneaux and P.~van Driel,
  Class.\ Quant.\ Grav.\  {\bf 12}, 2961 (1995)
  [arXiv:gr-qc/9506019].
\bibitem{Solodukhin:1998tc}
  S.~N.~Solodukhin,
  Phys.\ Lett.\  B {\bf 454}, 213 (1999)
  [arXiv:hep-th/9812056].
\bibitem{Cardy:1986ie}
  J.~L.~Cardy,
  Nucl.\ Phys.\  B {\bf 270} (1986) 186.\\
  H.~W.~J.~Bloete, J.~L.~Cardy and M.~P.~Nightingale,
  Phys.\ Rev.\ Lett.\  {\bf 56}, 742 (1986).
\bibitem{Carlip:1998wz}
  S.~Carlip,
  Phys.\ Rev.\ Lett.\  {\bf 82}, 2828 (1999)
  [arXiv:hep-th/9812013].
\bibitem{Carlip:1999cy}
  S.~Carlip,
  Class.\ Quant.\ Grav.\  {\bf 16}, 3327 (1999)
  [arXiv:gr-qc/9906126].
\bibitem{Lin:1999gf}
  F.~L.~Lin and Y.~S.~Wu,
  Phys.\ Lett.\  B {\bf 453}, 222 (1999)
  [arXiv:hep-th/9901147].\\
  V.~O.~Solovev,
  Phys.\ Rev.\  D {\bf 61}, 027502 (2000)
  [arXiv:hep-th/9905220].\\
  M.~Natsuume, T.~Okamura and M.~Sato,
  Phys.\ Rev.\  D {\bf 61}, 104005 (2000)
  [arXiv:hep-th/9910105].\\
  D.~J.~Navarro, J.~Navarro-Salas and P.~Navarro,
  Nucl.\ Phys.\  B {\bf 580}, 311 (2000)
  [arXiv:hep-th/9911091].\\
  J.~l.~Jing and M.~L.~Yan,
  Phys.\ Rev.\  D {\bf 62}, 104013 (2000)
  [arXiv:gr-qc/0004061].
\bibitem{Das:2000zs}
  S.~Das, A.~Ghosh and P.~Mitra,
  Phys.\ Rev.\  D {\bf 63}, 024023 (2001)
  [arXiv:hep-th/0005108].
\bibitem{Jing:2000yn}
  J.~l.~Jing and M.~L.~Yan,
  Phys.\ Rev.\  D {\bf 63}, 024003 (2001)
  [arXiv:gr-qc/0005105].\\
  H.~Terashima,
  Phys.\ Lett.\  B {\bf 499}, 229 (2001)
  [arXiv:hep-th/0011010].\\
  M.~Hotta, K.~Sasaki and T.~Sasaki,
  Class.\ Quant.\ Grav.\  {\bf 18}, 1823 (2001)
  [arXiv:gr-qc/0011043].
\bibitem{Terashima:2001gn}
  H.~Terashima,
  Phys.\ Rev.\  D {\bf 64}, 064016 (2001)
  [arXiv:hep-th/0102097].
\bibitem{Koga:2001vq}
  J.~i.~Koga,
  Phys.\ Rev.\  D {\bf 64}, 124012 (2001)
  [arXiv:gr-qc/0107096].
\bibitem{Park:2001zn}
  M.~I.~Park,
  Nucl.\ Phys.\  B {\bf 634}, 339 (2002)
  [arXiv:hep-th/0111224].
\bibitem{Carlip:2002be}
  S.~Carlip,
  Phys.\ Rev.\ Lett.\  {\bf 88}, 241301 (2002)
  [arXiv:gr-qc/0203001].
\bibitem{Cvitan:2002cs}
   M.~Cvitan, S.~Pallua and P.~Prester,
  Phys.\ Lett.\  B {\bf 555}, 248 (2003)
  [arXiv:hep-th/0212029].\\
  A.~J.~M.~Medved, D.~Martin and M.~Visser,
  Class.\ Quant.\ Grav.\  {\bf 21}, 3111 (2004)
  [arXiv:gr-qc/0402069].\\
  G.~Kang, J.~i.~Koga and M.~I.~Park,
  Phys.\ Rev.\  D {\bf 70}, 024005 (2004)
  [arXiv:hep-th/0402113].\\
  A.~Giacomini,
  Phys.\ Rev.\  D {\bf 70}, 044005 (2004)
  [arXiv:hep-th/0403219].\\
  M.~Cvitan, S.~Pallua and P.~Prester,
  Phys.\ Rev.\  D {\bf 70}, 084043 (2004)
  [arXiv:hep-th/0406186].\\
  S.~Carlip,
  Class.\ Quant.\ Grav.\  {\bf 22}, 1303 (2005)
  [arXiv:hep-th/0408123].\\
  M.~Cvitan and S.~Pallua,
  Phys.\ Rev.\  D {\bf 71}, 104032 (2005)
  [arXiv:hep-th/0412180].\\
  S.~Carlip,
  Class.\ Quant.\ Grav.\  {\bf 22}, R85 (2005)
  [arXiv:gr-qc/0503022].\\
  S.~Carlip,
  Phys.\ Rev.\ Lett.\  {\bf 99}, 021301 (2007)
  [arXiv:gr-qc/0702107].\\
  S.~Carlip,
  Gen.\ Rel.\ Grav.\  {\bf 39}, 1519 (2007)
  [Int.\ J.\ Mod.\ Phys.\  D {\bf 17}, 659 (2008)]
  [arXiv:0705.3024 [gr-qc]].\\
  L.~M.~Cao, Y.~Matsuo, T.~Tsukioka and C.~M.~Yoo,
  Phys.\ Lett.\  B {\bf 679}, 390 (2009)
  [arXiv:0906.2267 [hep-th]].\\
  L.~Rodriguez and T.~Yildirim,
  Class.\ Quant.\ Grav.\  {\bf 27}, 155003 (2010)
  [arXiv:1003.0026 [hep-th]].\\
  B.~K.~Button, L.~Rodriguez and C.~A.~Whiting,
  arXiv:1009.1661 [hep-th].\\
  H.~Chung,
  Phys.\ Rev.\  D {\bf 83}, 084017 (2011)
  [arXiv:1011.0623 [gr-qc]].\\
  A.~Guneratne, L.~Rodriguez, S.~Wickramasekara and T.~Yildirim,
  arXiv:1206.2261 [hep-th].
\bibitem{Park:1999tj}
  M.~I.~Park and J.~Ho,
  Phys.\ Rev.\ Lett.\  {\bf 83}, 5595 (1999)
  [arXiv:hep-th/9910158].\\
  S.~Carlip,
  Phys.\ Rev.\ Lett.\  {\bf 83}, 5596 (1999)
  [arXiv:hep-th/9910247].
\bibitem{Park:1999hs}
  M.~I.~Park and J.~H.~Yee,
  Phys.\ Rev.\  D {\bf 61}, 088501 (2000)
  [arXiv:hep-th/9910213].
 \bibitem{Dreyer:2001py}
  O.~Dreyer, A.~Ghosh and J.~Wisniewski,
  Class.\ Quant.\ Grav.\  {\bf 18}, 1929 (2001)
  [arXiv:hep-th/0101117].
\bibitem{Silva:2002jq}
  S.~Silva,
  Class.\ Quant.\ Grav.\  {\bf 19}, 3947 (2002)
  [arXiv:hep-th/0204179].
\bibitem{Brustein:2000fw}
  R.~Brustein,
  Phys.\ Rev.\ Lett.\  {\bf 86}, 576 (2001)
  [arXiv:hep-th/0005266].
\bibitem{Cvitan:2002rh}
 M.~Cvitan, S.~Pallua and P.~Prester,
  Phys.\ Lett.\  B {\bf 546}, 119 (2002)
  [arXiv:hep-th/0207265].
\bibitem{Giacomini:2003cg}
  A.~Giacomini and N.~Pinamonti,
  JHEP {\bf 0302}, 014 (2003)
  [arXiv:gr-qc/0301038].\\
  N.~Pinamonti and L.~Vanzo,
  Phys.\ Rev.\  D {\bf 69}, 084012 (2004)
  [arXiv:hep-th/0312065].
\bibitem{Dias:2006vy}
  G.~A.~S.~Dias and J.~P.~S.~Lemos,
  Phys.\ Rev.\  D {\bf 74}, 044024 (2006)
 [arXiv:hep-th/0602144].
\bibitem{Majhi:2011ws} 
  B.~R.~Majhi and T.~Padmanabhan,
  Phys.\ Rev.\ D {\bf 85}, 084040 (2012)
  [arXiv:1111.1809 [gr-qc]].\\
  B.~R.~Majhi and T.~Padmanabhan,
  arXiv:1204.1422 [gr-qc] (to appear in Phys. Rev. D (Rapid Comm.)).
\bibitem{Russo:1995zm} 
  J.~G.~Russo,
  Phys.\ Lett.\ B {\bf 359}, 69 (1995)
  [hep-th/9507009].
\bibitem{Deser:1976yx} 
  S.~Deser, M.~J.~Duff and C.~J.~Isham,
  Nucl.\ Phys.\ B {\bf 111}, 45 (1976).\\
  M.~J.~Duff,
  Nucl.\ Phys.\ B {\bf 125}, 334 (1977).
\bibitem{Polyakov:1981rd} 
  A.~M.~Polyakov,
  Phys.\ Lett.\ B {\bf 103}, 207 (1981).
\bibitem{Balbinot:1999vg}
  R.~Balbinot, A.~Fabbri and I.~L.~Shapiro,
  Nucl.\ Phys.\  B {\bf 559}, 301 (1999)
  [arXiv:hep-th/9904162].
\bibitem{Kolekar:2010py} 
  S.~Kolekar and T.~Padmanabhan,
  Phys.\ Rev.\ D {\bf 83}, 064034 (2011)
  [arXiv:1012.5421 [gr-qc]].
\bibitem{Leutwyler:1984nd}
  H.~Leutwyler,
  Phys.\ Lett.\  B {\bf 153}, 65 (1985)
  [Erratum-ibid.\  {\bf 155B}, 469 (1985)].
\end{thebibliography}
\end{document}